\documentclass[aps,prd,twocolumn,superscriptaddress, nofootinbib, notitlepage,longbibliography]{revtex4-1}  

\usepackage{amssymb,amsmath,amsfonts,amsbsy}
\usepackage{color}
\usepackage{enumerate}
\usepackage{graphicx}
\usepackage{pifont}

\usepackage{subfigure}

\usepackage{cancel}

\RequirePackage[colorlinks,citecolor=blue,urlcolor=blue,linkcolor=blue]{hyperref}

\usepackage[normalem]{ulem}

\begin{document}

\title{Testing gravity with cosmic variance-limited pulsar timing array correlations}

\author{Reginald Christian Bernardo}
\email{reginald.bernardo@apctp.org}
\affiliation{Asia Pacific Center for Theoretical Physics, Pohang 37673, Korea}
\affiliation{Institute of Physics, Academia Sinica, Taipei 11529, Taiwan}

\author{Kin-Wang Ng}
\email{nkw@phys.sinica.edu.tw}
\affiliation{Institute of Physics, Academia Sinica, Taipei 11529, Taiwan}
\affiliation{Institute of Astronomy and Astrophysics, Academia Sinica, Taipei 11529, Taiwan}

\begin{abstract}
The nanohertz stochastic gravitational wave background (SGWB) is an excellent early universe laboratory for testing the fundamental properties of gravity. In this letter, we elucidate on the full potential of pulsar timing array (PTA) by utilizing cosmic variance-limited, or rather experimental noise-free, correlation measurements to understand the SGWB and by extension gravity. We show that measurements of the angular power spectrum play a pivotal role in the PTA precision era for scientific inferencing. In particular, we illustrate that cosmic variance-limited measurements of the first few power spectrum multipoles enable us to clearly set apart general relativity from alternative theories of gravity. This ultimately conveys that PTAs can be most ambitious for testing gravity in the nanohertz GW regime by zeroing in on the power spectrum.
\end{abstract}

\maketitle

\textit{Introduction.---}Pulsar timing arrays (PTAs) have achieved an astronomical milestone \cite{NANOGrav:2023gor, Reardon:2023gzh, Antoniadis:2023lym, Xu:2023wog} by detecting the nanohertz stochastic gravitational wave background (SGWB) \cite{Burke-Spolaor:2018bvk, NANOGrav:2020spf}. This is done so through timing observation of millisecond pulsars (MSPs) whose radio pulses' time of arrival perturbations are spatially correlated by the SGWB \cite{Detweiler:1979wn}. This traditional revered signature is known as the Hellings-Downs (HD) correlation \cite{Hellings:1983fr, Romano:2016dpx}, that is dominantly quadrupolar and is shaped by the same gravitational wave (GW) polarizations that are now so familiar through ground-based GW detectors. The detection of the HD could be regarded as the `holy grail' of PTA science, and the present missions \cite{NANOGrav:2020bcs, Goncharov:2021oub, Chen:2021rqp, 2010CQGra..27h4013H} assembled under one banner `International PTA' (IPTA) have since been gearing up to resolve this elusive signal \cite{Allen:2023kib}.

\begin{figure}[h!]
    \centering    \includegraphics[width = 0.475\textwidth]{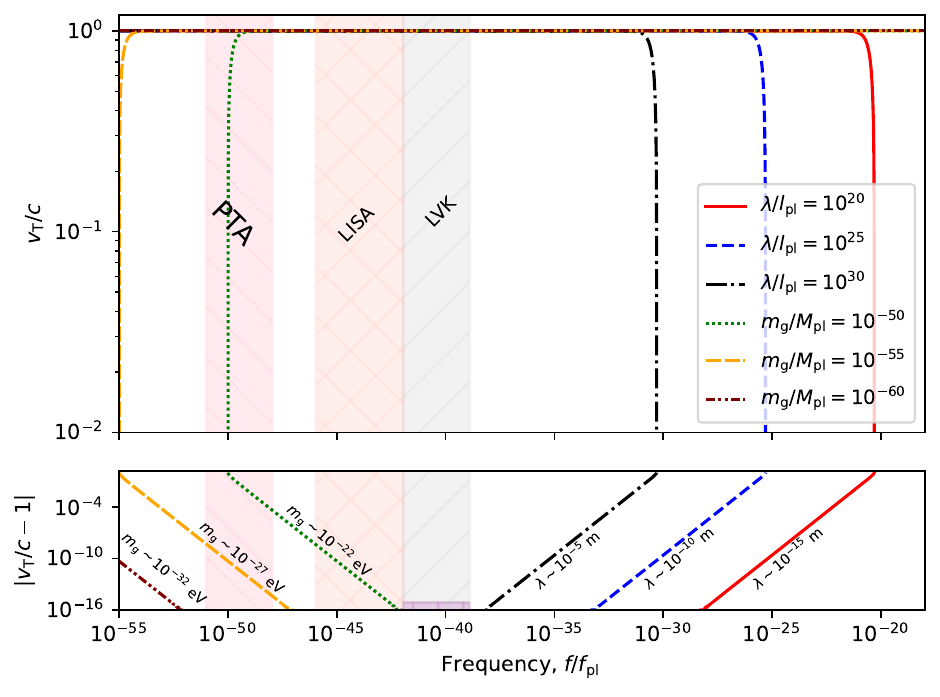}
    \caption{GW speed $v_{\rm T}$ frequency $f$ dependence in Horndeski (Eq. \eqref{eq:horndeski}) and massive gravity (Eq. \eqref{eq:massgrav}). The LIGO/Virgo/KAGRA (LVK) GW speed constraint translates to the Horndeski-Gauss-Bonnet coupling $\lambda \lesssim 10^{-5} \ {\rm m}$ and graviton mass $m_{\rm g} \lesssim 10^{-22} \ {\rm eV}{\mathbf{/c^2}}$.}
    \label{fig:vg}
\end{figure}

The scientific value of the HD cannot be understated, particularly when it is viewed as a byproduct of general relativity (GR) that is one in a multiverse of viable alternative theories of gravity (AG) \cite{Clifton:2011jh, Joyce:2014kja}. This becomes more meaningful in light of the multimessenger GW speed constraint, $|v_{\rm T}/c - 1| \lesssim 10^{-15}$, that has singlehandedly ruled out the parameter space of gravity that distorts the tensor modes' propagation cone \cite{Lombriser:2015sxa, Sakstein:2017xjx, Creminelli:2017sry, Baker:2017hug, Ezquiaga:2017ekz, Boran:2017rdn}. However, gravity and its attendant rainbow \cite{deRham:2018red} is also extraordinarily resilient to a few observational efforts and thus calls for a multiband GW objective where PTA's irreplaceable role in constraining gravity at low frequencies can be perfectly realized. This rainbow mechanism is portrayed in Fig.~\ref{fig:vg} for scalar-tensor (Eq. \eqref{eq:horndeski}) and massive gravity (Eq. \eqref{eq:massgrav}) that respect the stringent GW speed measurement in the subkilohertz GW band all while manifesting nontrivial GW propagation at different frequencies.

In this letter, we establish PTAs unparalleled potential in constraining gravity in the nanohertz GW regime through measurements of the angular power spectrum \cite{Gair:2014rwa, Qin:2018yhy,Qin:2020hfy,Ng:2021waj,Bernardo:2022rif,Liang:2023ary}. We consider cosmic variance-limited PTA measurements that can be expected to become relevant in future PTA missions when the experimental noise drops considerably below the cosmic variance of the GW correlation \cite{Allen:2022dzg, Allen:2022ksj,Bernardo:2022xzl}, marking the dawn of PTA precision era. As a template for gravity beyond GR, we consider widely-studied scalar-tensor theory and massive gravity to tease out non-Einsteinian GW polarizations that may inhabit the SGWB. Our codes and notebooks that derive all analytical and numerical results in this letter are publicly available in \href{https://github.com/reggiebernardo/PTAfast/tree/main/app3_cvlimitedgravity}{GitHub}. {In the following we consider natural units $c = \hbar = 1$.}

\medskip
\textit{Scalar-tensor and massive gravity.---}We consider the action functional \cite{Clifton:2011jh, Kase:2018aps}
\begin{equation}
\label{eq:horndeski}
\begin{split}
    S\left[g_{ab}, \phi\right] = & \int d^4x \sqrt{-g} \bigg[ \dfrac{M_{\rm pl}\left( M_{\rm pl} + \alpha \phi \right)}{2} R \\
    & + V\left(\phi, \left(\partial\phi\right)^2\right) + G\left(\phi, \left(\partial\phi\right)^2\right) \Box \phi + F(\phi) \hat{{\cal G}} \bigg] \,,
\end{split}
\end{equation}
where $M_{\rm pl}$ is the reduced Planck mass, $g_{ab}$ is the metric, $R$ is the Ricci scalar, $\hat{{\cal G}} = R^{abcd}R_{abcd} - 4 R^{ab}R_{ab} + R^2$ is the Gauss-Bonnet (GB) term, $\phi$ is a scalar field, $\alpha$ is a constant, and $V$, $G$, and $F$ are arbitrary potentials. This theory admits a static vacuum, $R_{ab} = 0$ and $\phi = m_v$, where the mass scale $m_v$ characterizes the scalar field's vacuum expectation value, provided that the potential $V$ meets the criterion $V\left(m_v, 0\right) = 0$ and $\partial V\left(m_v, 0\right)/\partial\phi = 0$. We take this static vacuum to represent the galactic weak field where GWs propagate.

We can find the propagating degrees of freedom (d.o.f.) and their causal structure in the perturbations. First off, we mention that the vector modes are nondynamical in a scalar-tensor theory. Now by massaging the constraint equations, it can be shown that the scalar d.o.f. satisfies the Klein-Gordon equation, $\ddot{\psi} - \partial^2 \psi + \mu^2 \psi = 0$, where
\begin{equation}
\label{eq:gal_mass}
\mu^2 = - \dfrac{\partial^2_\phi V\left(m_v, 0\right)}{\partial_X V\left(m_v, 0\right) + 2 \partial_\phi G\left(m_v, 0\right) + \epsilon } \, ,
\end{equation}
and $\epsilon = 3 M_{\rm pl} \alpha^2/\left(2 \left( 
M_{\rm pl} + m_v \alpha \right) \right)$. In the tachyonic instability-free region $\mu^2 > 0$, we focus on propagating scalar modes that satisfy the massive dispersion relation $\omega^2 = k^2 + \mu^2$, where $\omega = 2\pi f$ and $\vec{k} = k \hat{k}$ are the frequency and wave number, respectively. This leads to a group velocity $v_{\rm S} = d\omega/dk = k/\omega = 1/v_{\rm ph}$, where $v_{\rm ph}$ is the phase velocity. Putting these together, we find that the scalar d.o.f. projects scalar transverse (ST) and longitudinal (SL) metric polarizations \cite{Qin:2020hfy, Bernardo:2022vlj}: $h_{AB} \propto \varepsilon_{AB}^{\text{ST}} + \varepsilon_{AB}^{\text{SL}} \left(1-v_{\rm S}^2\right)/\sqrt{2}$, where $\varepsilon_{AB}^P$ are orthonormal polarization basis tensors. The longitudinal response can be attributed to a nonzero effective mass.

We can similarly show that the tensor d.o.f.s have a modified causual structure, depending on the coupling $\lambda$ which is given by
\begin{equation}
\label{eq:gaussbonnetlength}
    \lambda^2 = - \dfrac{8 F\left( m_v \right)}{M_{\rm pl}\left( M_{\rm pl} + m_v \alpha \right)} \,.
\end{equation}
In general, $\lambda \neq 0$, the tensor modes are pushed off the light cone. To keep the tensor modes from propagating superluminally, we consider $\lambda^2 > 0$, or in terms of the scalar field vacuum expectation values, $F(m_v)/(M_{\rm pl} + m_v \alpha) < 0$. We can tease out the dispersion structure of the tensor modes, $\omega^2 - k^2 + \lambda^2 k^4 = 0$, and thereby calculate its group velocity $v_{\rm T} = d\omega/dk$. This leads to the transverse GW spin-2 polarizations, $\varepsilon_{AB}^+$ and $\varepsilon_{AB}^\times$. This furthermore shows that the GB coupling acts as a friction that slows down the tensor modes upon approaching a cutoff frequency, $f \sim f_{\rm c} \simeq 0.5 (l_{\rm pl}/\lambda) f_{\rm pl}$, where $l_{\rm pl} \sim 10^{-35}$ m and $f_{\rm pl} \sim 10^{42}$ Hz are the Planck length and frequency, respectively. For order unity vacuum expectation values, $m_v \sim M_{\rm pl}$ and $F(m_v) \sim {\cal O}(1)$, the GB coupling $\lambda \sim \l_{\rm pl}$ and the cutoff frequency $f_{\rm c} \sim f_{\rm pl}$, implying that the tensor modes move at practically the speed of light. The parameter space $\lambda \lesssim 10^{-5} \ {\rm m}$ gives tensor GWs indistinguishable with GR\footnote{Other terms in the ${\cal L}_{4, 5}$ Horndeski couplings may give rise to a more relevant tensor sector \cite{Clifton:2011jh, Kase:2018aps}.} (Fig. \ref{fig:vg}). This motivates looking toward the scalar sector for scalar GWs' observable consequences.

Also, we consider massive gravity for a different brand of gravity, given by the Fierz-Pauli action \cite{Liang:2021bct}:
\begin{equation}
\label{eq:massgrav}
\begin{split}
    S[h] = & \int d^4 x \bigg[ \dfrac{\partial_c h_{ab} \partial^c h^{ab}}{2} - \partial_a h_{bc} \partial^b h^{ac} \\
    & + \partial_a h^{ab} \partial_b h - \dfrac{\partial_c h \partial^c h}{2} + \dfrac{m_{\rm g}^2}{2} \left( h_{ab} h^{ab} - h^2 \right) \bigg] \,,
\end{split}
\end{equation}
describing tensor, vector, and scalar modes with dispersion relation, $\omega^2 = k^2 + m_{\rm g}^2$, where $m_{\rm g}$ is the graviton mass. In contrast with Horndeski theory, the tensor modes' dispersion relation in massive gravity remain relevant in the PTA band for $m_{\rm g} \lesssim 10^{-22} \ {\rm eV}$ (Fig. \ref{fig:vg}). The theory further gives rise to vector excitations, which however are irrelevant and were shown to be disfavored \cite{Bernardo:2023mxc, Wu:2023pbt}. The scalar excitations of massive gravity are trivial due to Vainshtein screening \cite{Wu:2023pbt}. Nonetheless, for the discussion, we shall keep a complete set of modes representative of AG d.o.f.s in a PTA.

\medskip
\textit{Gravity in PTA.---}The SGWB spatially correlates the timing residuals of MSPs. This correlation traditionally expressed as $\gamma_{ab}(\zeta)$ \cite{Ng:2021waj, Bernardo:2022rif, Bernardo:2022xzl, NANOGrav:2021ini} between a pair of pulsars $a$ and $b$ separated by an angle $\zeta$ in the sky embodies the d.o.f.s of the superposed gravitational field that is in the SGWB, taking the form of the HD curve in GR. However, AG comes with d.o.f.s different from GR that can manifest with their distinct PTA correlation trademarks \cite{Chamberlin:2011ev}. Departures from the HD correlation thus tell which d.o.f.s live in the SGWB.

Demanding a precise understanding of the SGWB beyond GR is therefore only fair to meet such an ambitious science objective. The experimental demands are as high but the IPTA and future PTA iterations are prepared to handle the required precision. The future Square Kilometer Array (SKA) will cover a few thousand MSPs \cite{Janssen:2014dka, Weltman:2018zrl} and is forecasted to be at least three orders of magnitude more sensitive than existing PTAs.

The recent theoretical progresses on the SGWB have shed new insights on GW correlations that take over in PTAs. The spotlight goes to the harmonic analysis \cite{Gair:2014rwa,Qin:2018yhy,Qin:2020hfy,Ng:2021waj,Bernardo:2022rif,Liang:2023ary}, 
\begin{equation}
\label{eq:meancorr}
\gamma_{ab}\left(\zeta\right) = \sum_{l} \dfrac{2l+1}{4\pi} C_l P_l\left(\cos \zeta\right) \,,
\end{equation}
which recasts the correlation using the angular power spectrum, $C_l$'s, and the calculations of the variance of the HD curve \cite{Allen:2022dzg, Allen:2022ksj} and non-Einsteinian GW correlations \cite{Bernardo:2022xzl}, e.g., provided PTA correlations data from a sufficiently large number of pulsars, then the `Gaussian' nature of the SGWB manifests as the cosmic variance, $\Delta \gamma_{ab}\left(\zeta\right) \sim \sigma^{\rm CV}_{ab}\left(\zeta\right) \propto \sum_l (2l+1)C_l^2P_l\left(\cos\zeta\right)$, a natural limit to precision \cite{Ng:1997ez, Allen:2022dzg, Bernardo:2022vlj, Bernardo:2023bqx}. This influences mainly large scales and it is possible to identify theoretically when the experimental noise becomes subdominant, as in the cosmic microwave background (CMB).

Gravity in a PTA can thus be associated with the SGWB angular power spectrum, $C_l$'s, which further enables a fast and reliable numerical route to the correlations, and paves the road for the most general calculation of the correlations \cite{Bernardo:2022rif, Bernardo:2022xzl}. To put this plainly, all the past decades' work on the SGWB correlations can now be put together beautifully in {\it two lines and a table}:
\begin{equation}
\label{eq:powerspectrummultipoles}
    C_l^{\rm A} = {\cal F}_l^{\rm A}\left( fD,{v_{\rm g}} \right) {\cal F}_l^{\rm A}\left( fD,{v_{\rm g}} \right)^* / \sqrt{\pi}
\end{equation}
and
\begin{equation}
\label{eq:projectionfactors}
    \dfrac{{\cal F}_l^{\rm A}\left( y,{v_{\rm g}} \right)}{N_l^{\rm A}} = \int_0^{\frac{2\pi y}{v_{\rm ph}}} dx \ v_{\rm ph} e^{ixv_{\rm ph}} \dfrac{d^q}{dx^q} \left( \dfrac{j_l(x) + r v_{\rm ph}^{-2} j_l''(x)}{x^p} \right) \,,
\end{equation}
where $j_l(x)$'s are the spherical Bessel functions, $v_{\rm ph}$ is the phase velocity, {$v_{\rm g}$} is the GW speed {(group velocity)}, $f$ is the GW frequency, $D$ are the pulsar distances, and the coefficients $N_l^{\rm A}$ and indices $p, q, r$ are given in Table \ref{tab:powerspectrumcoefs}.

\begin{table}[h!]
\centering
\caption{The SGWB angular power spectrum coefficients and indices (Eqs. \eqref{eq:powerspectrummultipoles} and \eqref{eq:projectionfactors}) for arbitrary GW modes A.}
\label{tab:powerspectrumcoefs}
\begin{tabular}{|r|r|r|r|r|}
\hline
GW modes A & $N_l^{\rm A}/\left(2 \pi i^l\right)$ & \phantom{g} $p$ & \phantom{g} $q$ & \phantom{g} $r$ \\
\hline \hline
Tensor & \phantom{g} $\sqrt{(l+2)!/(l-2)!}/\sqrt{2}$ & 2 & 0 & 0 \\ \hline
Vector & $\sqrt{2} \sqrt{l(l + 1)}$ & 1 & 1 & 0 \\ \hline
Scalar & $1$ & 0 & 0 & 1 \\ \hline
\end{tabular}
\end{table}

We consider $v_{\rm g} = 1/v_{\rm ph}$ which holds for the tensor GWs in massive gravity and the scalar modes in Horndeski theory. But, we emphasize that other theories can have a different relation, $v_{\rm g} = v_{\rm g}\left(v_{\rm ph}\right)$, depending on their underlying dispersion relation \cite{Liang:2023ary}.

The ${\cal F}^{\rm A}_l(y, v{_{\rm g}})$'s are generalization to the projection factors given astrophysical pulsar distances \cite{Qin:2018yhy, Qin:2020hfy}. For scalar GWs, it is worth noting that the integration reduces to a vanishing boundary term\footnote{{For details, see supplementary `Scalar modes' boundary integration'.}} for $l \geq 2$ as $fD\rightarrow \infty$ \cite{Qin:2020hfy, Bernardo:2022vlj}. This shows that scalar GWs can be distinguished by their pronounced monopolar and dipolar powers. On the other hand, tensor and vector GWs are noteworthy for their significant quadrupolar and dipolar powers, respectively.

Now we show that this recipe empowers us to draw the line between GR and AG using PTA data.

\medskip
\textit{Variance in the power spectrum.---}By appealing to the Gaussianity of the SGWB, we can show that the variance of the power spectrum is given by~\cite{Allen:2022dzg,Bernardo:2022xzl}
\begin{equation}
\label{eq:clcv}
    \left( \dfrac{\Delta C_l}{C_l} \right)^2 = \dfrac{2}{2l + 1} \,,
\end{equation}
reminiscent of the CMB. We can expect that in a few years the PTAs will be able to resolve low $C_l$'s ($l \lesssim 8$) with concealed experimental noise. The recently forecasted 30 years PTA observation of 150 MSPs \cite{Nay:2023pwu} shows that the first few multipoles can be resolved to a subfraction of the cosmic variance (Eq. \eqref{eq:clcv}). Even the projected 10 years PTA observation of only 50 MSPs \cite{Nay:2023pwu} shows experimental noises that are comparable to the cosmic variance. This is supported by the NANOGrav 15 years data \cite{NANOGrav:2023gor}, e.g., Fig. 7 of Ref. \cite{NANOGrav:2023gor} shows experimental noise smaller than the cosmic variance up to $l \leq 5$. This optimism translates to the heart of this letter--that cosmic variance-limited measurements of the SGWB angular power spectrum is the key to understanding gravity in the nanohertz GW regime. More lightly, cosmic variance resolution of the first few power spectrum multipoles is sufficient so that PTAs can test gravity.

\begin{figure}[h!]
    \centering
    \begin{minipage}[b]{0.45\textwidth}
        \centering
        \includegraphics[width=\textwidth]{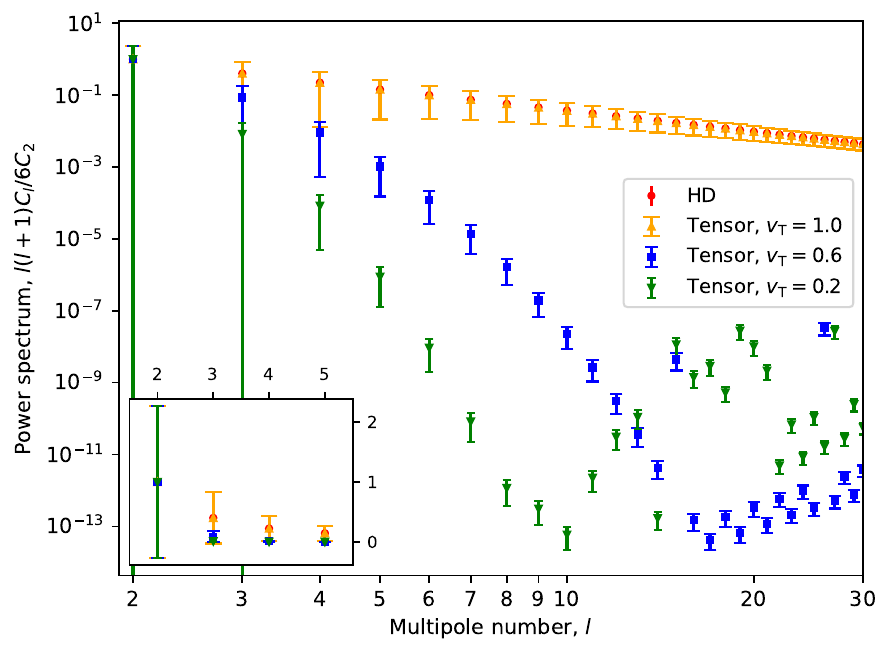}
    \end{minipage}\hfill
    \begin{minipage}[b]{0.45\textwidth}
        \centering
        \includegraphics[width=\textwidth]{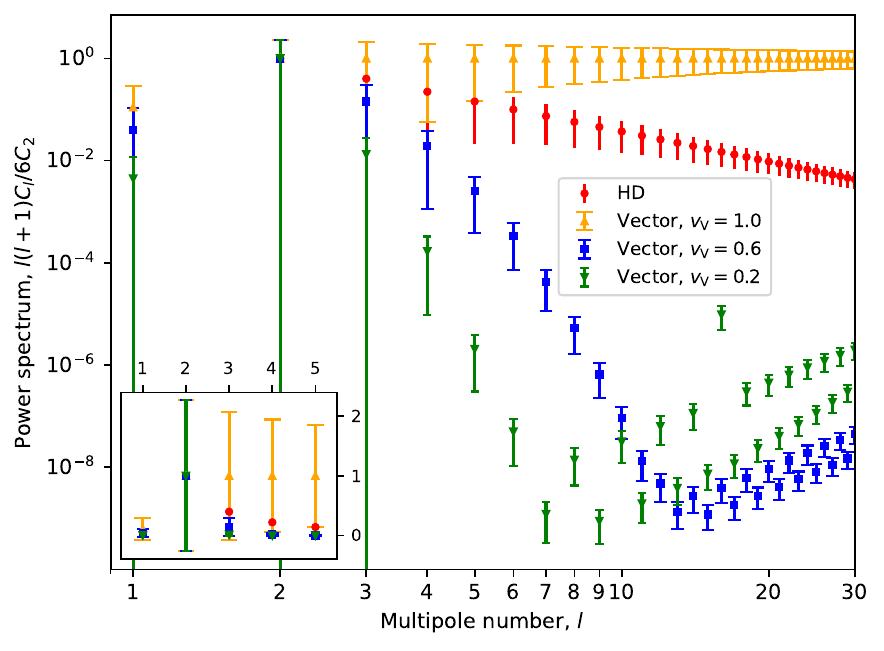}
    \end{minipage}\hfill
    \begin{minipage}[b]{0.45\textwidth}
        \centering
        \includegraphics[width=\textwidth]{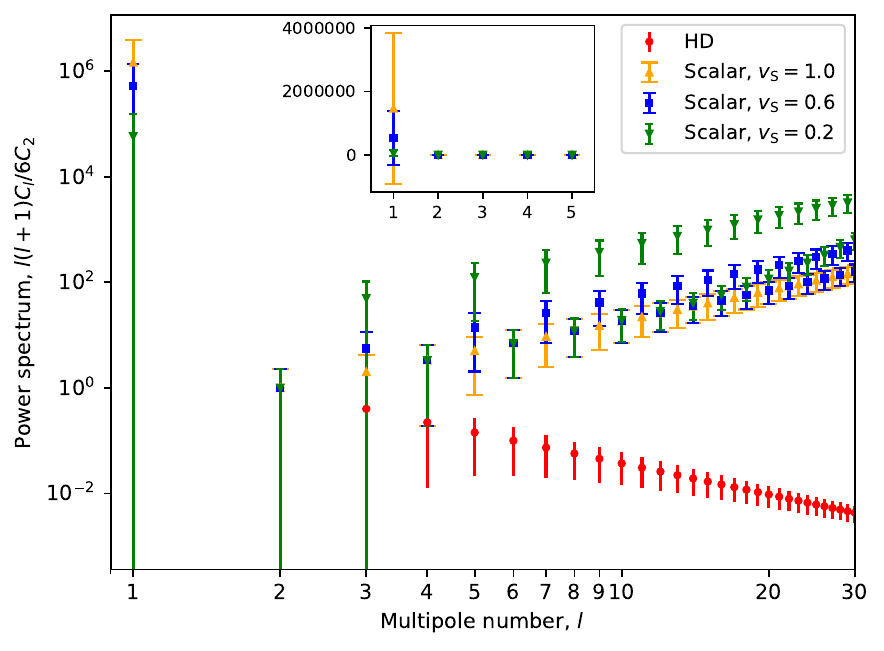}
    \end{minipage}
    \caption{The SGWB angular power spectrum ($C_l \pm 2\Delta C_l^{\rm CV}$) for tensor, vector, and the scalar modes. For scalar modes, $C_0/C_1 \sim 9, 25, 225$ for $v_{\rm S} \sim 1.0, 0.6, 0.2$, respectively.}
    \label{fig:powerspectrum}
\end{figure}

Figure \ref{fig:powerspectrum} strengthens this point, revealing the power spectrum of tensor, vector, and scalar GWs at various speeds, at a frequency $f \sim 1 \ {\rm yr}^{-1}$ with pulsars at a distance $D \sim 0.2 \ {\rm kpc}$. To use this for instance, we can associate subluminal tensor GWs to massive gravity (Eq. \eqref{eq:massgrav}) and scalar modes to Horndeski theory (Eq. \eqref{eq:horndeski}). Then, it becomes clear that the cosmic variance-limited SGWB power spectrum measurements can be used to constrain the GW speed, even with only low $l \lesssim {\cal O}(10)$'s. Higher multipoles, $l \gtrsim {\cal O}(10)$, of subluminal GWs are susceptible to environment contamination, e.g., finite distance effects \cite{Ng:2021waj, Bernardo:2022rif}, in addition to the experimental challenges of resolving them. To be explicit, we consider massive gravity, giving tensor GWs with a speed determined by the graviton mass, $m_{\rm g}$, i.e., $v_{\rm T} = v_{\rm T}(m_{\rm g})$. Cosmic variance-limited measurements of $C_l$ thus give away $v_{\rm T}$, which then exposes $m_{\rm g}$. Interpreted together with LVK's stringent bound on the coupling (Fig. \ref{fig:vg}), then subluminal GW propagation in PTA would be unequivocal evidence for the long sought after graviton with a mass $m_{\rm g} \sim 10^{-22} \ {\rm eV}$. In the same vein, if the data expresses monopolar and dipolar powers unaccounted by systematics, then this can be associated scalar d.o.f.s. The ratios between the monopole, dipole, and quadrupole can be used to consistently determine the speed of scalar GWs, $v_{\rm S} = v_{\rm S}\left(\mu\right)$, which traces back to the coupling constants of the theory and the field potentials.

\begin{figure}[h!]
    \centering    \includegraphics[width = 0.475\textwidth]{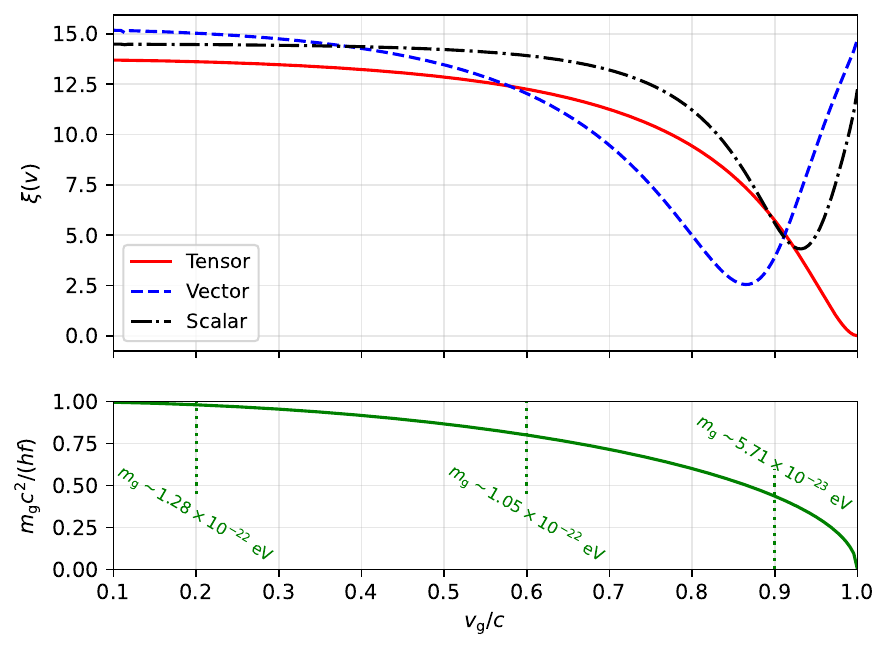}
    \caption{(top) Eq. \eqref{eq:xidef} with $L = 5$ for tensor, vector, and scalar modes with GW speeds $v_{\rm g}$. (bottom) Ratios $m_{\rm g}/f$ corresponding to the GW speeds $v_{\rm g}$. Vertical dotted labels show the graviton masses at a reference frequency $f \sim 1 \ {\rm yr}^{-1}$.}
    \label{fig:xi}
\end{figure}

The departures from the HD correlation (or GR) must be put in a solid grounding, in terms of the power spectrum, in order to put forthcoming PTA data to best use. A {possible} metric is 
\begin{equation}
\label{eq:xidef}
    \xi(v{_{\rm g}}) = \sum_{l \leq L} \dfrac{\left( C_l - \overline{C}_l \right)^2}{ \Delta C_l^2 + \Delta \overline{C}_l^{2} } = \sum_{l \leq L} \dfrac{2l+1}{2}\dfrac{\left( C_l - \overline{C}_l \right)^2}{ C_l^2 + \overline{C}_l^2 } \,,
\end{equation}
where $\overline{C}_l$ and $\Delta \overline{C}_l$ are the corresponding HD multipoles and uncertainty from the cosmic variance. Eq. \eqref{eq:xidef} reads the deviation of a measured power spectrum, $C_l \pm \Delta C_l$ with $\Delta C_l \sim C_l \sqrt{2/\left(2l+1\right)}$, from the HD. Figure \ref{fig:powerspectrum} suggests that we evaluate $\xi(v{_{\rm g}})$ only until environmental effects set in, and this depends on how fast we expect GWs propagate. A modest upper limit is the quintupole as shown in Fig.~\ref{fig:xi}.

\begin{figure}[h!]
    \centering    \includegraphics[width = 0.475\textwidth]{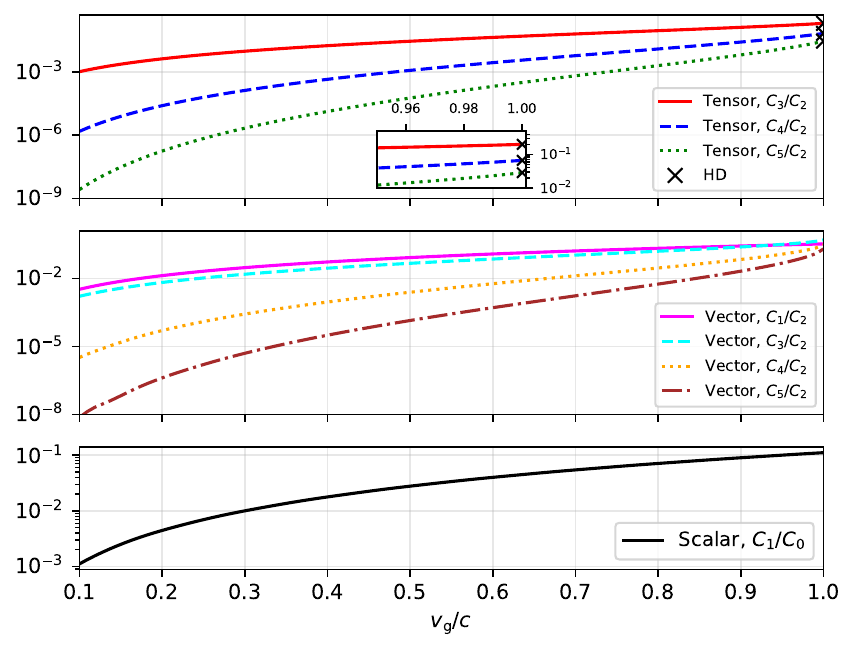}
    \caption{Power spectrum ratios for tensor, vector, and scalar modes weighted by their dominant multipole ($C_2$ for tensor and vector, $C_0$ for scalar).}
    \label{fig:clsratios}
\end{figure}

This shows HD deviations, in a form that PTAs can utilize for fundamental physics, e.g., if there is negligible monopolar and dipolar powers and $\xi \sim 5-6$, then this implies $m_{\rm g} \sim 5.71 \times 10^{-23} \ {\rm eV}$ and that the SGWB is dominantly tensor-natured and quadrupolar, the natural hypothesis. Figure \ref{fig:xi} also shows that the extra monopolar and dipolar powers in the scalar and vector modes causes large deviations away from the HD at near luminal GW speeds, $v_{\rm g} \sim 0.9-1.0$ and reaches a minimum $\xi(v_{\rm min})$ when higher multipoles begin to notably drop away from the HD. The pronounced deviation of vector and scalar modes at $v_{\rm g} \sim c$ is inherited from their divergence in the luminal GW speed and infinite distance limit \cite{NANOGrav:2021ini, Qin:2020hfy, Bernardo:2022rif}. However, in practice, it may happen that the dominant multipole will be correlated with the estimated GW amplitude, thus compromising the power spectrum's magnitude \cite{Nay:2023pwu}. In this case, we can appeal to power spectrum ratios with respect to the dominant multipole for science inferencing (Fig. \ref{fig:clsratios}), utilizing the $C_l$'s ratios tailored values depending on the GW speed. An important comment goes to the scalar modes where the nonvanishing of $C_l$'s for $l \geq 2$ can be distinctly associated with environmental factors and so $C_1/C_0$ seems to be the singularly relevant factor.

Either way, it is transparent that the power spectrum give away the nature of the SGWB, and that PTAs will be in the best position to test gravity by focusing on the power spectrum.

\medskip
\textit{Outlook.---}The playbook setup in this letter can be tested with the NANOGrav 15 years data \cite{NANOGrav:2023gor}, giving $\xi \sim 20$ for $l \geq 2$, which at face value suggests subluminal GW propagation having suppressed octupole and higher multipoles. However, this can be largely explained by the negative power spectrum estimates, unsupported by relativistic d.o.f.s. The nontrivial monopolar and dipolar powers could furthermore be associated with scalar modes, although the error bands are too wide to constrain their speed, i.e., $C_1/C_0 \sim 0-1$. Granted, the experimental noise remain nonnegligible even when it is now subdominant to the cosmic variance \cite{NANOGrav:2023gor}. This crude application nonetheless attests that we are on track.

Our results are not without caveats. We were assuming that the pulsars are on a sphere, while the reality is that MSPs are at different distances. Nonetheless, the takeaway is that finite distance effects, which we artificially associate environmental factors, do not significantly influence the low multipoles, $l \lesssim 10$, of relativistic d.o.f.s in the SGWB. This gives reason to utilize the first few power spectrum multipoles for scientific inferencing with PTA. The cosmic variance is also anchored on a Gaussianity hypothesis of the SGWB, which remains to be tested, and astronomical factors such as there is a large number of evenly distributed pulsars. Future PTA and SKA measurements can be expected to overcome these and test the Gaussianity of the SGWB.

It is worthwhile to also consider the possible mixture of tensor, vector, and scalar GWs in the SGWB \cite{Bernardo:2023zna}, which is relevant for astrophysical sources \cite{Callister:2017ocg}. These different modes are excited by very different processes, and since the SGWB hosts an exciting variety of high energy physics phenomena, looking out for the non-Einsteinian polarizations is a reasonable way to go to maximize science objectives.

PTAs are lastly part of a multiband array of GW telescopes, and so interpreting the results from the larger GW picture should ultimately be the way forward. The SGWB can be viewed as the tip of the iceberg for PTA science since behind this galactic GW superposition may lie hints of a universe that is otherwise hidden from astronomical probes such as the CMB. This letter through the power spectrum highlights PTA science's invaluable edge in GW astronomy moving forward. 

\medskip

We thank Meng-Xiang Lin for important comments over a preliminary draft of this paper. This work was supported in part by the National Science and Technology Council (NSTC) of Taiwan, Republic of China, under Grant No. MOST 111-2112-M-001-065. RCB is supported by an appointment to the JRG Program at the APCTP through the Science and Technology Promotion Fund and Lottery Fund of the Korean Government, and was also supported by the Korean Local Governments in Gyeongsangbuk-do Province and Pohang City.


%

\clearpage
\appendix

\widetext

\section*{Testing gravity with cosmic variance-limited \\ pulsar timing array correlations: Supplementary notes}

\setcounter{equation}{0}
\setcounter{page}{1}

\section*{In and out of natural units}
\label{sec:natural_units}

Natural units, $c = \hbar = 1$, is an often considered system in high energy physics to make the symmetries among various physical variables be more transparent, e.g., $E = mc^2$ becomes $E = m$, making more explicit the relation between the energy and the rest mass, and the massive dispersion relation $\hbar^2 \omega^2 = \hbar^2 c^2 k^2 + m^2 c^4$ simplifies to $\omega^2 = k^2 + m^2$. In natural units, velocity is dimensionless; length and time are measured in the same units; energy, momentum, and mass have the same units; length and energy units have inverse relations to each other. To go in and out of natural units requires simply substituting back the appropriate factors of $c$ and $\hbar$ to make the units right. Several examples are given below.

Take the values $c \simeq 3 \times 10^8 \ {\rm m/s}$ and $h = 2\pi \hbar \simeq 4.136 \times 10^{-15} \ {\rm eV \cdot s}$. Then setting $c = \hbar = 1$ makes the conversion factors between length/time and energy/time to be
\begin{equation}
    3 \times 10^{8} \ {\rm m} \simeq 1 \ {\rm s}
\end{equation}
and
\begin{equation}
    4.136 \times 10^{-15} / (2\pi) \ {\rm eV} \simeq 1 \ {\rm s}^{-1} \,.
\end{equation}
This lets one translate mass, energy, and momentum to electron volts (eV) and length and time to inverse electron volts (eV$^{-1}$). One meter is $\simeq 2\pi \times 8.06 \times 10^5 \ {\rm eV}^{-1}$. The way out of natural units back to the traditional SI system is paved by the same conversion factors.

In natural units, velocities are also dimensionless, and happen to be their ratios with the speed of light in vacuum. The relation, $v_{\rm g} = 1/v_{\rm ph}$, between the group and phase velocity derived with a massive dispersion relation makes sense since $v_{\rm g}$ and $v_{\rm ph}$ are dimensionless; whereas in the traditional system the relation becomes $v_{\rm g} = c^2/v_{\rm ph}$.

\section*{Scalar modes' boundary integration}
\label{sec:scalarmodesboundaryintegration}

For scalar GWs, the dominant multipoles are the monopole and the dipole; the quadrupole and higher multipoles are subdominant compared with these first two. This is shown analytically by the following integration.

The two core principles that go to show this are wave superposition and partial integration. Starting with the relation for scalar GWs,
\begin{equation}
    h_{AB} \propto \varepsilon_{AB}^{\text{ST}} + \varepsilon_{AB}^{\text{SL}} \left(1-v_{\rm S}^2\right)/\sqrt{2} \,,
\end{equation}
which can be derived from a general scalar-tensor action, the relevant scalar GWs' projection factor can be written as
\begin{equation}
    {\cal F}^{\rm S}_l \left(fD\right) \propto - \int_0^{2\pi fDv_{\rm S}} \dfrac{dx}{v} e^{ix/v_{\rm S}} \left( R^{\rm ST}_l(x) + \dfrac{1-v_{\rm S}^2}{\sqrt{2}} R^{\rm SL}_l(x)  \right) \,,
\end{equation}
where the $R_l(x)$'s can be found in \cite{Qin:2020hfy, Bernardo:2022rif}, i.e., $R_l^\text{SL}(x) = j_l''(x)$ and $R_l^\text{ST}(x) = -\left( R_l^\text{SL}(x) + j_l(x) \right) / \sqrt{2}$. Then the total scalar field-induced projection factor becomes 
\begin{equation}
\begin{split}
    F^{\rm S}_l \left(fD\right) \propto & - \int_0^{2\pi fDv_{\rm S}} \dfrac{dx}{v_{\rm S}} e^{ix/v_{\rm S}} \left( -\dfrac{j_l''(x) + j_l(x)}{\sqrt{2}} + \dfrac{1-v_{\rm S}^2}{\sqrt{2}} j_l''(x)  \right) \\
    \propto & - \int_0^{2\pi fDv_{\rm S}} \dfrac{dx}{v_{\rm S}}  \dfrac{e^{ix/v_{\rm S}}}{\sqrt{2}} \left( -j_l(x) - v_{\rm S}^2 j_l''(x)  \right) \,.
\end{split}
\end{equation}
This is the integral expression that comes up when using the simplified formula in the letter that takes the insight that the ST and SL modes do not come separately in a relativistic theory but rather only simultaneously as a combination depending on the scalar modes' speed. This consequently leads to a destructive interference/suppression of the total higher-mode (e.g., quadrupole, octupole, and so on) contribution to the correlation. By noting that
\begin{equation}
    e^{ix/v_{\rm S}} j_l''(x) = \dfrac{d}{dx}\left(e^{ix/v_{\rm S}} j_l'(x)\right) - \dfrac{d}{dx}\left( \dfrac{i}{v_{\rm S}}e^{ix/v_{\rm S}}j_l(x) \right) - \dfrac{e^{ix/v_{\rm S}}}{v_{\rm S}^2} j_l(x) \,,
\end{equation}
then the projection factor can be seen to reduce to a couple of boundary terms,
\begin{equation}
    F^{\rm S}_l \left(fD\right) \propto - \dfrac{1}{\sqrt{2}} \int_0^{2\pi fDv_{\rm S}} \dfrac{dx}{v_{\rm S}} \left[ \dfrac{d}{dx} \left( iv_{\rm S} e^{ix/v_{\rm S}} j_l(x) \right) - \dfrac{d}{dx} \left( v_{\rm S}^2 e^{ix/v_{\rm S}} j_l'(x) \right) \right] \,.
\end{equation}
For $l \geq 2$, the lower limits of this integral vanishes, thus giving the final analytical result
\begin{equation}
    F^{\rm S}_l \left(fD\right) \propto - \dfrac{e^{2\pi ifD}}{\sqrt{2}v_{\rm S}} \left[ iv_{\rm S} j_l(2\pi f Dv_{\rm S})  -  v_{\rm S}^2 j_l'(2\pi fDv_{\rm S})  \right] \,.
\end{equation}
In the infinite pulsar distance limit, $fD \rightarrow \infty$, this vanishes, and therefore reduces to the result of \cite{Qin:2020hfy}. On the other hand, for finite, astrophysical pulsar distances, which was one of our novelty, clearly these terms remain but remain subdominant compared with the monopole and the dipole.

The takeaway of the calculation is that scalar GWs are revealed by pronounced monopolar and dipolar powers with a specific ratio that is related to the scalar modes' propagation speed. This is true whether in the infinite distance limit or when environmental factors set in such as the finite distances of the pulsars. This was first argued for $f(R)$ gravity in \cite{Qin:2020hfy} and generalized in the letter to a much wider class of scalar-tensor theory. \medskip

\end{document}